\documentclass[twocolumn,showpacs,preprintnumbers,nofootinbib,amssymb]{revtex4-1}

\usepackage[dvipdfm]{graphicx} 
\usepackage{epsfig, amssymb} 
\usepackage{amsmath, amsfonts}
\usepackage{bm} 

\usepackage[linktocpage]{hyperref}
\usepackage[caption=false]{subfig}
\usepackage[usenames]{color}

\DeclareMathOperator\erf{erf}

\newcommand{\dif}{{\rm d}}

\newcommand{\tA}{\tilde{A}}

\def\Lie{\mathcal{L}}

\def\H{\mathcal{H}}
\def\M{\mathcal{M}}

\def\beq{\begin{equation}}
\def\eeq{\end{equation}}
\def\beqa{\begin{eqnarray}}
\def\eeqa{\end{eqnarray}}

\def\p{\partial}

\def\tA{\tilde{A}}

\begin{document}

\title{\large Black holes and fundamental fields:\\
hair, kicks and a gravitational ``Magnus'' effect}

\author{Hirotada Okawa}
\affiliation{CENTRA, Departamento de F\'{\i}sica, Instituto Superior T\'ecnico, Universidade de Lisboa,
Avenida Rovisco Pais 1, 1049 Lisboa, Portugal}

\author{Vitor Cardoso} 
\affiliation{CENTRA, Departamento de F\'{\i}sica, Instituto Superior T\'ecnico, Universidade de Lisboa,
Avenida Rovisco Pais 1, 1049 Lisboa, Portugal}
\affiliation{Perimeter Institute for Theoretical Physics Waterloo, Ontario N2J 2W9, Canada}
\affiliation{Department of Physics and Astronomy, The University of Mississippi, University, MS 38677, USA}


\begin{abstract} 
Scalar fields pervade theoretical physics and are a fundamental ingredient to solve the dark matter problem, 
to realize the Peccei-Quinn mechanism in QCD or the string-axiverse scenario. They are also a useful proxy
for more complex matter interactions, such as accretion disks or matter in extreme conditions.
Here, we study the collision between scalar ``clouds'' and rotating black holes.
For the first time we are able to compare analytic estimates and strong field, nonlinear numerical calculations for this problem.
As the black hole pierces through the cloud it accretes according to the Bondi-Hoyle prediction, but is deflected through a purely kinematic gravitational ``anti-Magnus'' effect, which we predict to be present also during the interaction of black holes with accretion disks. 
After the interaction is over, we find large recoil velocities in the transverse direction. 
The end-state of the process belongs to the vacuum Kerr family if the scalar is massless, but can be a hairy black hole when the fundamental scalar is massive.
\end{abstract}

\pacs{
04.70.-s
11.25.Wx,
14.80.Va,
98.80.Es,
04.30.-w,
}
\maketitle

\noindent{\bf{\em I. Introduction.}}
Black holes (BHs) are known to be abundant objects in our universe, 
with a major role in the evolution of galaxies and star formation.
As truly relativistic objects, they are powerful sources of gravitational waves
and key-players in the nascent field of gravitational wave astronomy. 
Astrophysical observations of BHs will give us unprecedented information about our universe,
by mapping the BH mass and spin with exquisite precision~\cite{McClintock:2011zq,Antoniadis:2013pzd,Reynolds:2013qqa,Berti:2009kk};
by testing General Relativity in the strong-field regime \cite{Gair:2012nm,Yunes:2013dva,Berti:2009kk};
by constraining the dark-energy equation of state \cite{Arun:2007hu}; or by providing information on dark matter distribution
around BHs \cite{Eda:2013gg,Macedo:2013qea}. 

Supermassive BHs also have the unexpected ability to provide information
on fundamental ultra-light bosonic degrees of freedom, generic predictions of beyond the standard model physics and of modified gravity theories~\cite{Peccei:1977hh,Arvanitaki:2010sy,Cardoso:2013fwa}. 
For boson masses in the range $10^{-21}{\rm eV}\lesssim \mu_S \lesssim 10^{-8}{\rm eV}$, 
the Compton wavelength of these fields is of the order of the BH size, the gravitational coupling of these two objects is strongest, 
and long-lived quasi-bound states arise~\cite{Cardoso:2005vk,Dolan:2007mj,Witek:2012tr}. 
Depending on the efficiency with which the bosonic cloud is accreted, one might observe gravitational wave ``light houses'' or find gaps in the BH-Regge plane~\cite{Arvanitaki:2010sy,Kodama:2011zc,Cardoso:2011xi,Witek:2012tr,Cardoso:2013krh,Pani:2012vp}.
The end-state of the superradiant instability is not known, but the prospect of finding long-lived --or even truly stationary--
``hairy'' BH solutions deserves all the attention possible~\cite{Okawa:2014nda,Hod:2012px,Herdeiro:2014goa}.

Fundamental fields are also a useful proxy for more complex interactions and matter. In this context, the interaction between BHs and bosonic fields
can teach us about BH formation from gravitational collapse, interaction with accretion disks, magnetic fields, etc.
The rich phenomenology of such natural theories prompted a flurry of activity in the field,
mostly confined to the \textit{linearized} regime where the spacetime is a fixed Kerr BH background.
The purpose of this {\em Letter} is to take the first step towards understanding the nonlinear development 
of the interaction between BHs and fundamental fields.
Unless stated otherwise, we use geometrical units $G=c=1$.

\noindent{\bf{\em II. Numerical Setup and Analysis Tools.}}
We consider a minimally coupled, gravitating, complex scalar field $\Phi$ of mass $\mu_S$ described by the action
\begin{equation}
S=\int d^4x\sqrt{-g} \left( \frac{\,^{(4)}R}{16\pi} 
        -\frac{1}{2} \p^{\mu}\Phi^{\ast}{}\p_{\mu}\Phi-\frac{\mu_S^2}{2}\Phi^{\ast}{}\Phi \right) 
\,, 
\end{equation}
where $\,^{(4)}R$ is the four-dimensional Ricci scalar. 
We employ standard Numerical Relativity techniques 
based on the $3+1$ splitting to solve the fully nonlinear problem~\cite{Alcubierre:2008,Baumgarte2010}.
In this approach the time evolution of the $3$-metric $\gamma_{ij}$ and scalar field $\Phi$ are governed by
\begin{align}
\label{eq:evolGamPhi}
\dif_t \gamma_{ij} = - 2\alpha K_{ij}
\,,\qquad&
\dif_t \Phi = - \alpha \Pi
\,,
\end{align}
where the extrinsic curvature $K_{ij}$ and $\Pi$ are their conjugated momenta, 
and $\dif_t = \p_{t} -\Lie_{\beta}$.
The $3+1$ decomposition of the equations of motion yields 
time evolution equations for the extrinsic curvature and scalar field momentum
\begin{subequations}
\label{eq:EvolEqADM}
\begin{align}
\label{eq:EvolKADM}
\dif_t K_{ij} = & - D_{i} D_{j} \alpha + \alpha\left( R_{ij} -2 K^{k}{}_{i} K_{jk} + K K_{ij} \right) \nonumber\\
        &+ 4\pi\alpha\left(\gamma_{ij}(S-\rho) - 2S_{ij} \right)\,,\\
\label{eq:EvolKphiADM}
\dif_t \Pi = & 
        \alpha\left( -D^{i}D_{i}\Phi + K \Pi + \mu_S^2 \Phi\right)
        - D^{i}\alpha D_{i} \Phi 
\,,
\end{align}
\end{subequations}
as well as the Hamiltonian and momentum constraints
\begin{subequations}
 \label{eq:constraintsADM}
  \begin{align}
   \label{eq:HamiltonianADM}
   \H = & R + K^2 - K_{ij} K^{ij} - 16\pi \rho = 0 
   \,,\\
   \label{eq:MomentumADM}
   \M_{i} = & D_{j} K^{j}{}_{i} - D_{i} K - 8 \pi j_{i} = 0
   \,,
  \end{align}
\end{subequations}
where $R_{ij}$ and $R$ are associated to the $3$-dimensional metric 
and $\rho, j_{i}, S_{ij}$ are, respectively, the energy density, energy-momentum flux and 
spatial components of the energy momentum tensor.
In practice, we employ the Baumgarte-Shapiro-Shibata-Nakamura formulation~\cite{Baumgarte:1998te,Shibata:1995we}
of the evolution equations~\eqref{eq:evolGamPhi} and~\eqref{eq:EvolEqADM}, together 
with the moving puncture gauge~\cite{Campanelli:2005dd,Baker:2005vv}.

We solve the Cauchy problem for Einstein's equations using the COSMOS code~\cite{Okawa:2014nda}.
Time evolution is realized by a 4th order Runge-Kutta method,
spatial derivatives are computed through a 4th order finite
differencing method in Cartesian grids. The Adaptive Mesh Refinement algorithm of moving boxes is employed
in order to keep a good resolution near the BH~\cite{Berger:1984zza,Bruegmann:2006at,Yamamoto:2008js}.
Apparent horizons are tracked using the methods outlined in Refs.~\cite{Shibata:1997nc,Shibata:2000nw}.
Parallelization is implemented with OpenMP.

We measure the scalar field amplitude $\Phi$ and the Newman-Penrose scalar $\Psi_4$ encoding the gravitational radiation,
at coordinate spheres of fixed radius $r_{\rm{ex}}$, where we project them with
spherical or $s=-2$ spin-weighted spherical harmonics. We estimate the numerical discretization error to be of order $\lesssim6\%$ in both the scalar and gravitational waveforms.
In addition, we monitor the apparent horizon (AH) area $A_{\rm{AH}}$, the irreducible mass and the ratio of equatorial to polar circumferences to
estimate the BH mass and spin \cite{Kiuchi:2009jt,Sperhake:2009jz}.

\noindent{\bf{\em III. Initial data construction.}}
In general, one needs appropriate initial data to perform reliable and realistic
simulations within numerical relativity.
Following the initial data construction in Refs.\cite{Cook:2000vr,Okawa:2014nda},
we simplify the constraint equations by the conformal transformation
\begin{align}
\label{eq:conformaltrafoID}
\gamma_{ij} = \psi^4\eta_{ij} \quad {\rm and}\quad
K_{ij} \equiv \psi^{-2}\tA_{ij} +\frac{1}{3}\gamma_{ij}K\,,
\end{align}
where $\eta_{ij}$ is the flat metric.
Assuming conformal and asymptotic flatness, the maximal slicing condition and setting scalar field $\Phi(t=0)=0$,
the constraints~\eqref{eq:constraintsADM} become 
\begin{subequations}
\label{eq:ConstraintsID}
\begin{align}
\mathcal{H} 
\label{eq:ham_const_t}
= & \bigtriangleup_{\rm flat}\psi 
 +\frac{1}{8}\tA^{ij}\tA_{ij}\psi^{-7}
 +\pi\psi^5\Pi^{2}=0\,,\\
 \mathcal{M}_i 
 \label{eq:mom_const_t}
 =& \partial_j\tA^j_i=0\,.
\end{align}
\end{subequations}
%

Let us consider first a single, non-rotating BH $(\tA_{ij}=0)$ and a nonzero scalar field.
The momentum constraints~\eqref{eq:mom_const_t} are trivially satisfied and the Hamiltonian constraint~\eqref{eq:ham_const_t} yields
\begin{align}
\label{eq:ham_const_a}
\bigtriangleup_{\rm flat}\psi
+\pi\Pi^2\psi^{5}
= & 0\,.
\end{align}
Using the same ansatz for the Gaussian-type spherical scalar wave packet
described in Ref.~\cite{Okawa:2014nda},
\begin{align}
\label{eq:ansatz_SBH}
 \Pi =& \frac{A_0}{2\pi}e^{-\frac{r^2}{w^2}}\psi^{-\frac{5}{2}}
 \,\quad {\rm and}\quad
 \psi = 1 +\frac{M_0}{2r_{BH}} +\frac{u_0(r)}{\sqrt{4\pi}r}
 \,,
\end{align}
where we take the radial coordinate $r=\sqrt{x^2+y^2+z^2}$, the location of BH is described by $r_{BH}\equiv\sqrt{(x-x_{BH})^2 +y^2 +z^2}$
and $M_0$ denotes the BH bare mass parameter. A regular, analytical, solution to the Hamiltonian constraint is then
%
%
%
\begin{align}
 \label{eq:u0}
 u_0= & \frac{A_0^2w^3}{16\sqrt{2}}
 \erf{\left(\frac{\sqrt{2}r}{w}\right)}
 \,,
\end{align}
where we have imposed that $u_{0}\to 0$ at $r=0$. Thus, eqs.\eqref{eq:ansatz_SBH}-\eqref{eq:u0} describe a spherically symmetric scalar cloud and a BH a distance $x_{BH}$ apart.


Addition of linear and angular momenta complicates the procedure, but can be done as follows.
The momentum constraints~\eqref{eq:mom_const_t} can be also solved
analytically and we obtain
the so-called Bowen-York extrinsic curvature
\begin{eqnarray}
 \label{eq:BYansatz}
 \tA_{BY}^{ij} &=& \frac{3}{2r^2}\left(P^{i} n^{j} + P^{j} n^{i} - (\eta^{ij} - n^{i} n^{j})P^{k} n_k \right)\nonumber \\
 &+& \frac{3}{r^3} \left( \epsilon^{ikl} S_{k} n_{l} n^{j} + \epsilon^{jkl}S_{k} n_{l} n^{i}\right)\,,
\end{eqnarray}
where $P^i, S_{i}$ and $n^i$ are the momentum, the spin
and the unit normal vector $n^i\equiv x^i/r$, respectively.
The remaining Hamiltonian constraint is then given by
\begin{align}
 \label{eq:HamPuncBH}
 \mathcal{H} = & \bigtriangleup_{\rm flat}\psi 
 + \frac{1}{8}\tA_{BY}^{ij}\tA^{BY}_{ij}\psi^{-7} 
 + \pi\psi^{5} \Pi^2 
 = 0
 \,.
\end{align}
To solve the Hamiltonian constraint, we use the ansatz
\begin{align}
 \label{eq:ansatz_conf_PBH}
 \psi = 1 +\frac{M_0}{2r} +u(x^i)\,,
\end{align}
where $M_0$ is the BH bare mass parameter and $u(x^i)$ is a regular function.
We set a boosted, rotating BH initially at the origin and
a scalar pulse located at $x=x_0$ apart from the BH.
The scalar cloud is again described by the Gaussian profile
\begin{align}
 \Pi = \frac{A_P}{2\pi}e^{-\frac{r_0^2}{w^2}}\psi^{-3}.
\end{align}
where $A_P$ and $w$ denote the amplitude and width of the scalar cloud
and $r_0=\sqrt{(x-x_0)^2+y^2+z^2}$.
Eq.~\eqref{eq:HamPuncBH} becomes an elliptic partial differential equation for $u$
which is regular everywhere
and can be solved with a common elliptic-equation solver~\cite{Okawa:2013afa}.
\begin{align}
 \label{eq:HamPuncBHu}
 \bigtriangleup_{\rm flat} u &=
 - \frac{1}{8}\tA_{BY}^{ij}\tA^{BY}_{ij}\psi^{-7} 
 - \frac{A_P^2}{4\pi}e^{-\frac{2r_0^2}{w^2}}\psi^{-1}
 \,.
\end{align}
%

\noindent{\bf{\em III. Collision of BHs with scalar ``clouds''.}} 
%
\begin{figure}[ht]
 \psfig{file=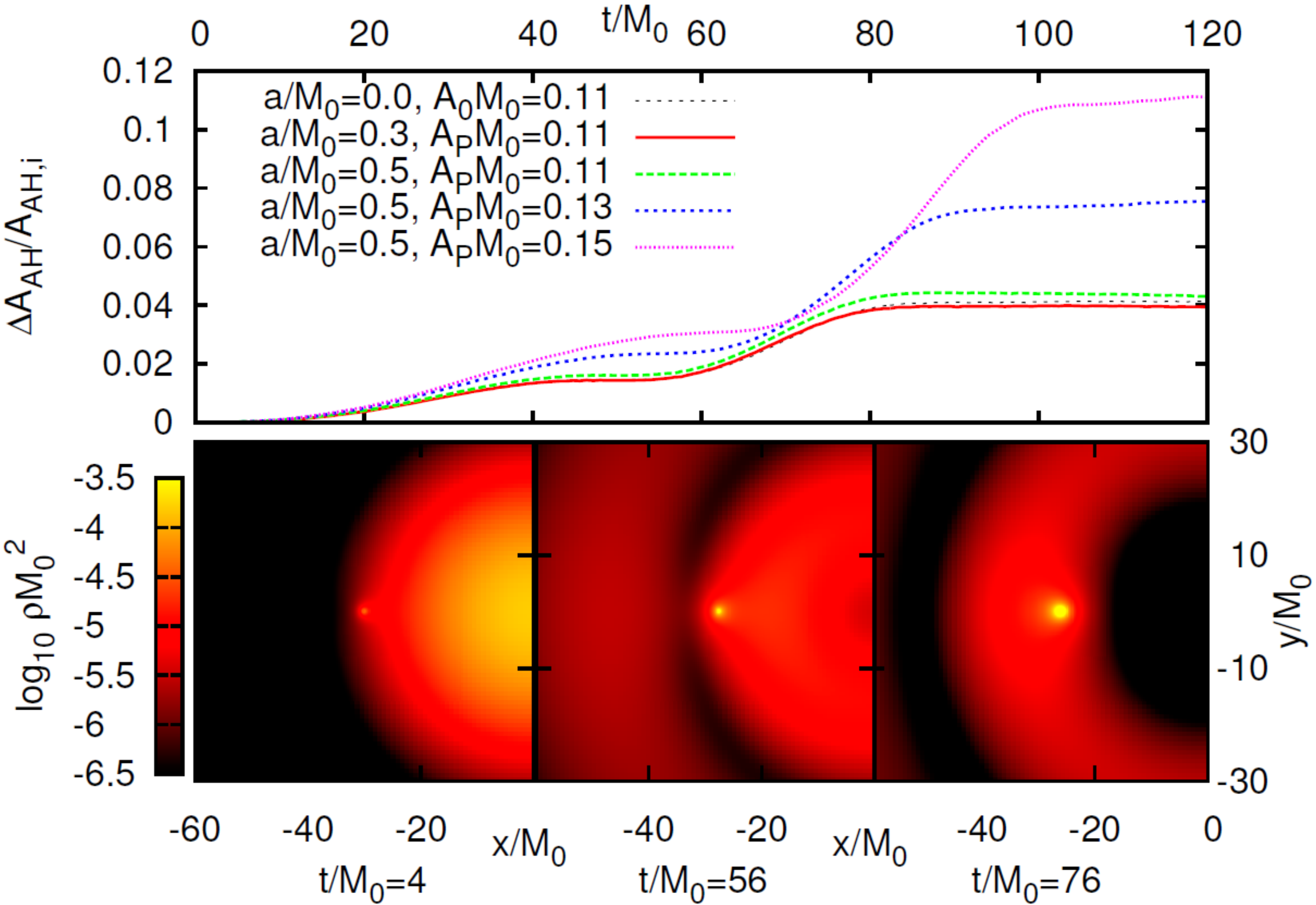,width=9cm,clip}
 \caption[]{{\it Upper panel:} Scalar field accretion. BH area increases for different
 initial BH spin and scalar-field amplitude. 
 The fundamental scalar is taken to be massless, and
 the scalar ``cloud'' is initial described by a Gaussian
 with $w=20M_0$ located at the origin for non-rotating BH and
 $x_0=30M_0$ for rotating BH.
 {\it Lower panel:} Snapshots of scalar-density on the $x-y$ plane at different instants, for $a/M_0=0.0$ and $A_0M_0=0.11$.
 The temporal density distribution explains the different accretions stages in the upper panel.
 }
 \label{fig:accretion}
\end{figure}
We have evolved a variety of different initial configurations, varying the BH mass momentum and spin, and varying the scalar-field width, location
and mass $\mu_S$. The collision process is gravity-dominated, and we find that timescales are well approximated by Newtonian free-fall estimates.
The evolution proceeds in different stages, depending on the scalar-field mass.

Consider the massless or small $\mu_S M$ regime first. For low scalar-field amplitudes, a fraction of the initial scalar cloud is unbounded and scatters to infinity. As we increase the Gaussian amplitude, we find that the scalar field starts self-gravitating and a larger fraction is accreted by the BH. These features are summarized in Fig.~\ref{fig:accretion}, specialized to a width $w=20 M_0$, which we take as representative for the rest of this work. For this setup, typically $99\%$ of the scalar-field mass escapes to infinity: the ``cloud'' is initially located far away from the BH
and is dispersing away from it. Our results are in quantitative agreement with Bondi-Hoyle accretion rates, they depend only weakly on BH spin, but scale like the square of the scalar field amplitude, as expected. There are two pronounced accretion phases, related to scalar cloud evolution and its density profile, as shown in the lower panel of Fig.~\ref{fig:accretion}. The final state is a Kerr BH in vacuum.

\begin{figure}[h]
 \includegraphics[width=9.0cm,clip]{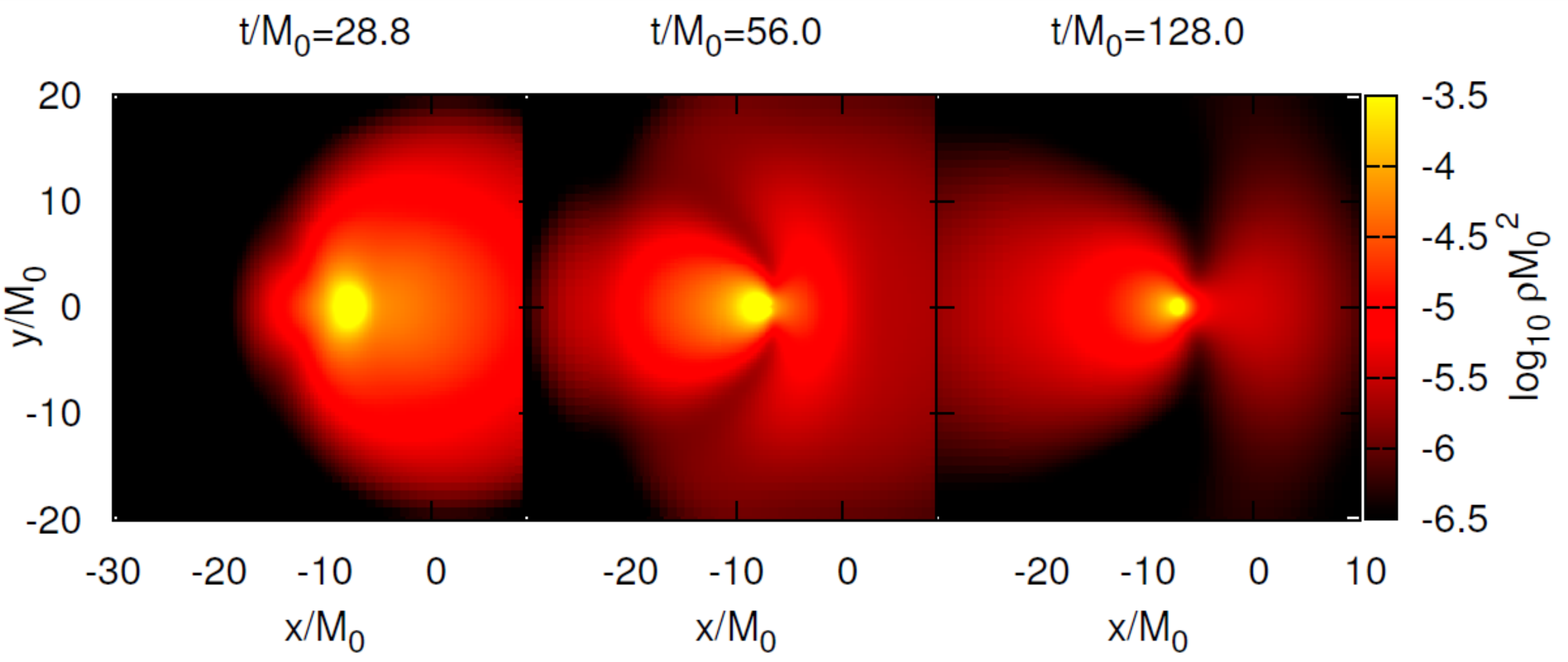} 
 \caption{Snapshots of the accretion flow to a non-rotating BH on $x-y$ plane.
 The field is initially described by a gaussian with $A_0M_0=0.2,\, x_{BH}=-8M_0,\, w=6M_0$ and has a mass term $\mu_S M_0=0.4$.
 Colors depict intensity of the scalar field, at late-times the configuration settles to a very long-lived
 dipole condensate outside the horizon, extending to distances of order $\sim 20M_0$~\cite{Witek:2012tr}.\label{fig:rho_a00}}
\end{figure}
The accretion and evolution of configurations where the fundamental scalar is massive shows qualitatively different behavior.
Massive fields are harder to disperse and try to bound. Accordingly we find substantial more amount of scalar field being accreted onto the BH
in the massive-scalar case. An intriguing alternative to this scenario is that -- if the BH is rotating--
superradiance prevents absorption of the scalar at the horizon and instead forms a {\it hairy} BH, with a non-trivial external scalar field configuration and quadrupole moment~\cite{Herdeiro:2014goa,Hod:2012px}. Even in the absence of rotation, extremely long-lived modes have been shown to
be possible~\cite{Cardoso:2005vk,Dolan:2007mj,Witek:2012tr,Okawa:2014nda}. Our results point to a possible formation scenario: a cloud of scalar field scattering off a non-rotating BH leaves behind, at late times, a BH surrounded by an external long-lived scalar condensate.
Snapshots of the evolution for a scalar with $\mu_S M_0=0.4$ are shown in Fig.~\ref{fig:rho_a00}. The pattern oscillates with a frequency compatible with linearized calculations which also describe well the spatial extent of the scalar condensate. In other words, we have strong evidence of a possible mechanism for formation of what for practical purposes is a ``hairy'' BH in asymptotically flat spacetime.

\noindent{\bf{\em IV. The (anti-)``Magnus'' effect in BH physics.}} 
As the BH pierces through the cloud, accretion of matter ensues. It is well-known that the absorption cross-section for co- and counter-rotating particles and waves is different for spinning BHs~\cite{Bardeen:1972fi,Futterman:1988ni}, causing a {\it kinematic} drift of general-relativistic origin in the perpendicular direction to the flow. Consider the BH initially at the center of the reference frame,
spinning with angular momentum $J$ aligned with the $z-$axis and moving in the $x-$direction through the scalar field cloud. The BH accretes mass as it moves, in a spin-dependent manner. For low-velocity collisions, accretion is governed by the marginally bound circular orbit of radius (in ``Brill-Lindquist'' coordinates~\cite{Bardeen:1972fi})
\begin{equation}
R_{\mp}=2M\mp a+2\sqrt{M^2\mp aM}\,.
\end{equation}
The upper (lower) sign applies to co-(counter-) rotating orbits.
Modeled in this way then, as the BH moves through the medium with relative velocity $v$, it sweeps up a distorted, non-symmetric ``tube''
composed by two half cylinders with radii $R_{-},\,R_+$. Finding the shape of this distorted tube is an interesting geometrical problem, which in its simplest version amounts to equating the centroid of the projected figure to the BH location, a problem similar in many respects to that found in two-dimensional rocket motion~\cite{1979rpsd.book.....C}.
A simple estimate can be obtained by noting that when the BH moves a distance $\delta x$, the $y-$position of the center-of-mass of this distorted cylinder is located at $\sim 2\rho(R_2^3-R_1^3)\delta x/3M$, with $\rho$ the energy density of the scalar configuration. Thus, after accretion of the material the BH has to sit in the CM, at $\delta y\gtrsim 2\rho v (R_{-}^3-R_{+}^3)\,\delta x/3M
\sim 100 M a \rho \delta_x$. 
\begin{figure}[ht]
 \psfig{file=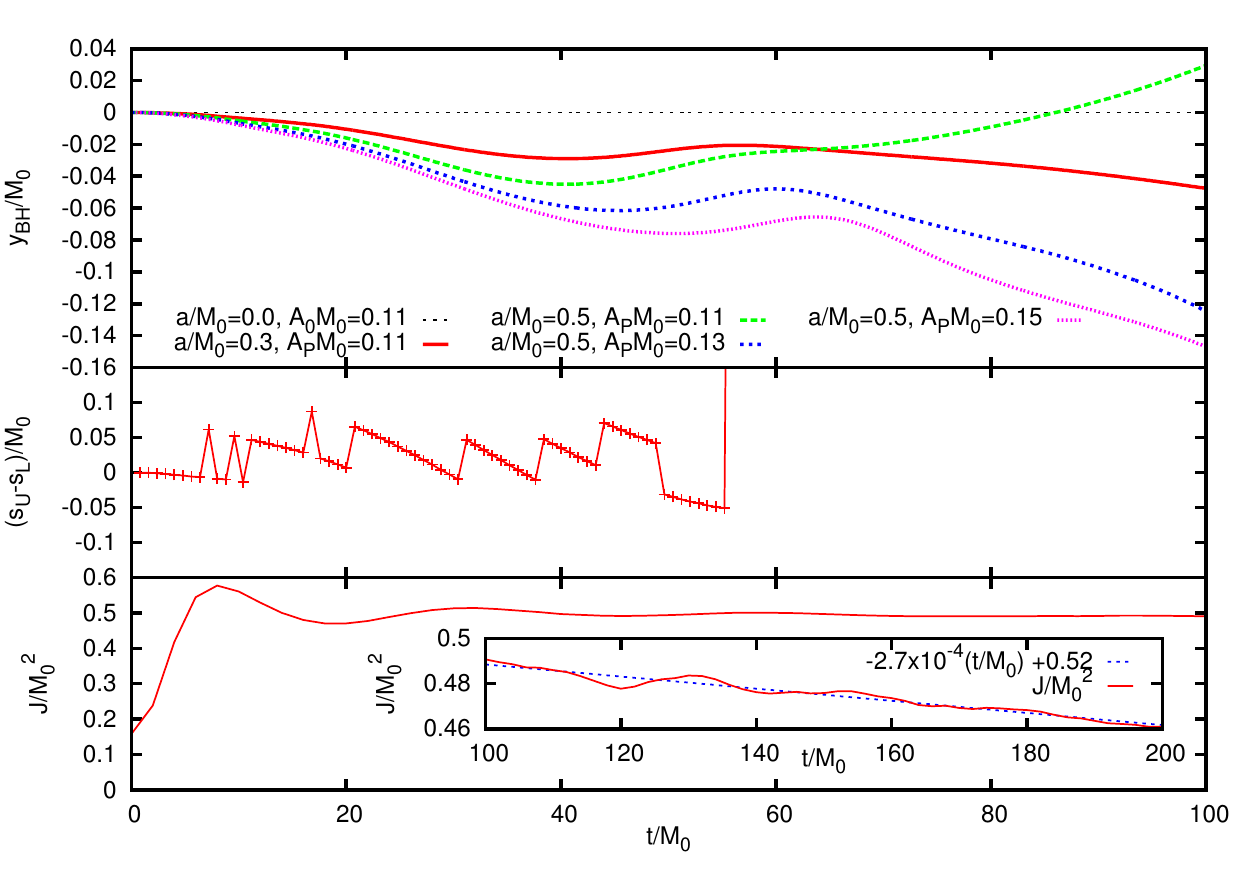,width=9cm,clip}
 \caption[color online]{The gravitational ``anti-Magnus'' effect.
 {\it Upper panel:} BH (puncture) $y-$coordinate as a function of time, showing two significant stages where the BH moves downwards. These two stages are consistent with the accretion pattern of Fig.~\ref{fig:accretion}.
 {\it Middle panel:} Difference between proper length from the BH
 to the upper and lower edge of the scalar cloud in the $y$-direction, for $a/M_0=0.5, A_PM_0=0.15$.
 The overall pattern is that of a {\it relative downward} movement of the BH relative to the cloud. 
 {\it Lower panel:} Total BH angular momentum $J$ for $a/M_0=0.5, A_PM_0=0.15$.
 It decreases at late times. 
 }
\label{fig:Magnus2}
\end{figure}
This new effect in BH physics, triggered by asymmetric accretion, is responsible for the motion in the direction orthogonal to the initial BH velocity.
In this respect, it is similar to the ``Magnus effect'' in fluid dynamics, a well-known corollary of hydrodynamics
with important consequences in sports, aeronautics, etc~\cite{Seifert:2012aa}\footnote{Magnus-like effects were also reported for Abrikosov vortices~\cite{Papanicolaou:1990sg} and an ``optical'' Magnus effect was predicted (and observed) to exist as well, by Zeldovich and collaborators~\cite{Zeldovich:1992a,Zeldovich:1992b}; similarities with general-relativistic equations of motion were put forward in Ref.~\cite{Duval:2005ky}. The effect we describe here is no analogy, it is a purely General Relativistic effect.}. However, (i) the original Magnus effect is a consequence of delicate boundary-layer effects close to the body's surface, whereas the BH drift we described is a pure consequence of spacetime drag and kinematics and (ii) the Magnus effect results, generically but not always, in a motion in the $y-$direction but in the {\it opposite} sense to the BH drift that we predict via spacetime drag and kinematics. 

Asymmetric accretion is potentially concurrent with three other effects present in our simulations.
The first is an overall momentum in the transverse direction triggered by scalar or gravitational waves,
potentially displacing the entire BH+cloud system. The second competing effect is the frame-dragging of the scalar cloud,
which again by momentum conservation would rigidly rotate the system. Both effects could mask the asymmetric accretion
deflection. However, we find that whereas the asymmetric accretion is expected to scale with the scalar cloud density (for a fixed total mass say),
this is no longer the case for the other two competing effects. Finally, nonhomogeneous media would give rise to asymmetric accretion
and a consequent transversal motion which would be rotation independent. In our setup the BH lies along the symmetry axis and such effect is non-existent.

Our numerical results are summarized in Fig.~\ref{fig:Magnus2}. The upper panel shows the puncture position along the $y$-axis as a function of time. These results are gauge-dependent and a simple overall coordinate shift in the negative $y$-direction would masque it. We have therefore also measured the proper distance in the $y$-direction from the BH to the outer boundaries of the cloud, defined as the points for which the density decreases to $1\%$ of its central value. If the BH really moves downwards {\it with respect to the cloud}, then the distance to the upper part of the cloud should be larger than the distance to the lower part of the cloud. This is in fact the overall pattern seen in the middle panel of Fig.~\ref{fig:Magnus2}.
Thus, an overall shift of the system does not explain our numerical results. The lower panel of Fig.~\ref{fig:Magnus2} shows the time evolution of the total angular momentum $J$ of the BH. In line with the predicted
preferred absorption of counter-rotating particles, $J$ decreases. Finally, our results are proportional to density, spin and velocity, as expected for the asymmetric accretion scenario we propose. 

This effect is not a particularity of scalar fields, but a rather general feature of BH interaction with matter.
An order of magnitude estimate for astrophysically realistic sources is given by
%
%
%
\beq
v_y 
\sim 10^{-2} \frac{f_{\rm Edd}^{11/20}}{\tilde{r}^{15/8}}\,  \left(\frac{0.1}{\alpha}\right)^{7/10} \left(\frac{M}{10^8 M_\odot}\right)^{13/10} \frac{a}{M}\,,\nonumber 
\eeq
where we take as reference value the density of thin accretion disks close to supermassive BHs, $f_{\rm Edd}$ is the Eddington luminosity, $\alpha$
is the disk's viscosity parameter and $\tilde{r}\equiv GM r/c^2$ is the distance of the BH from the center of the disk~\cite{shakura_sunyaev,2002apa..book.....F}.

These numbers are encouraging, and open-up the possibility to actually observe the gravitational anti-Magnus effect.
This deflection is all the more interesting as it can in addition provide an evidence for the existence of horizons:
compact stars or any other object with a surface will presumably be subjected to an ordinary Magnus effect.

We also observe large ``kicks'' in the transverse direction after the BH ceased interacting with the scalar cloud.
These kicks, presumably imparted by gravitational waves, are already apparent in the puncture position. Their magnitude depends 
on the scalar cloud amplitude and width. Further exploration of this effect is necessary to understand whether it is a viable
recoil mechanism in realistic astrophysical scenarios.

\noindent{\bf{\em V. Conclusions.}} 
We reported on the first steps towards understanding the interaction between fundamental fields and BHs.
Much remains to be understood, but we think our setup will be useful in exploring fundamental issues such
as fully nonlinear investigations of gravitational drag, turbulent wakes, spin alignment and spin precession during the interaction
of BHs with matter.

{\bf \em Acknowledgements.}
We are indebted to C. Herdeiro and P. Pani for useful comments and suggestions.
H. O. and V.C. acknowledge financial support provided under the European Union's FP7 ERC Starting Grant ``The dynamics of black holes:
testing the limits of Einstein's theory'' grant agreement no. DyBHo--256667.
This research was supported in part by Perimeter Institute for Theoretical Physics. 
Research at Perimeter Institute is supported by the Government of Canada through 
Industry Canada and by the Province of Ontario through the Ministry of Economic Development 
$\&$ Innovation.
This work was supported by the NRHEP 295189 FP7-PEOPLE-2011-IRSES Grant.
Computations were performed on the ``Baltasar Sete-Sois'' cluster at IST.

\bibliographystyle{h-physrev4}
\bibliography{Accretion}

\end{document}